\documentclass[12pt]{article}
\usepackage[dvips]{graphicx}
\usepackage[hang]{subfigure}

\newcommand{\bfi}{\begin{figure}}
\newcommand{\efi}{\end{figure}}
\newcommand{\subf}{\subfigure}
\newcommand{\bmi}{\begin{minipage}}
\newcommand{\emi}{\end{minipage}}

\begin{document}
\begin{center}
{\large\bf 
Neutrino-Induced Giant Air Showers in Large Extra Dimension Models
}



\bigskip
{\large\bf Ambar Jain$^a$, Pankaj Jain$^a$, Douglas W. McKay$^b$\\ and John P. Ralston$^b$}

\bigskip
$^a$Physics Department, I.I.T. Kanpur, India 208016\\
$^b$Department of Physics and Astronomy,\\ University of Kansas,\\ Lawrence,
KS 66045, USA\\ 
\end{center}

\bigskip

\noindent {\bf Abstract:}  In models based on large extra dimensions
where massive spin 2 exchange can dominate at high energies, the 
neutrino-proton 
cross section can rise to typical hadronic values at energies above
$10^{20}$ eV.  The neutrino then becomes a candidate for the primary
that initiates the highest energy cosmic ray showers.  We investigate
characteristics of neutrino-induced showers compared to
proton-induced showers. The comparison includes study of starting depth,
profile with depth, lateral particle distribution at ground and muon
lateral distribution at ground level. We find that for
cross sections above 20 mb there are regions of parameter space
where the two types of showers are essentially indistinguishable.  We
conclude that the neutrino candidate hypothesis cannot be ruled out
on the basis of shower characteristics.

\section{Introduction}
Many ultra high energy (UHE) cosmic ray air showers with energies in
excess of $5\times 10^{19}$ eV have been observed in the past few 
decades \cite{nw}. The nature and origin of the primary particles is not
understood \cite{nw,es,sigl}.  The puzzle is that
the sources have to be within the GZK limit of approximately 50 Mpc if
these particles are protons or nuclei \cite{GZK,puget}.  
However there are not enough
powerful astrophysical sources within this distance to explain the events.

Among known particles, only neutrinos travel larger distances than protons
in intergalactic space. This leaves neutrinos as the only established
candidates that can travel the  distances greater than 100 Mpc from known
UHE sources.  The GZK bound of 50 Mpc is not applicable to them. Yet neutrino
interactions with matter are too weak in the Standard Model of particle
physics to generate the observed air showers. Hence these events seem to demand
a revision of our current understanding of nature. Either the determination
of the number of sources of such ultrahigh energy particles
in our astrophysical neighborhood is grossly low\footnote{For example 
if magnetic fields outside the galaxy have been underestimated, ``line 
of sight'' and ``photon travel time'' requirements on protons and nuclei 
can be relaxed and new source possibilities considered \cite{es}},
or these observations are a signal of new physics.
  
Many speculative ideas have been proposed to explain the events above 
$10^{19} - 10^{20}eV$, including topological defects such as cosmic
strings \cite{abs,mb} and associated decays of heavy, relic particles
\cite{berez,kt}, existence of neutral, stable, strongly interacting
particles, such as a light gluino \cite{ch,raby,lou} or a monopole
bound state \cite{kw,wick}, and violation of Lorentz invariance
\cite{g-m,cg}.  Much of this work requires that the primary particle
responsible for generating these air showers is an exotic new particle
which does not exist within the Standard Model. 

In a recent paper \cite{us1} we argued that the data is consistent with
the general features of massive spin-2 exchange.  Models where effects of
gravity can be strong just above the weak scale
\cite{add} supply a natural and attractive framework.  The interaction cross
section of neutrinos with matter is greatly enhanced with massive spin-2
exchange at UHE and may reach values close to the hadronic cross sections.
In the low scale gravity models, the cross section enhancement arises from
t-channel exchange of the tower of gravitons.  Our estimates of the 
neutrino-proton cross section at the highest energies 
relevant for these events are
of the order of one to a few hundred millibarns.  
The highest energy cosmic ray events may therefore
be initiated by neutrinos\footnote{Correlation
between the positions of compact radio quasars and the track directions of
UHE $>100$ EeV cosmic rays has been studied by several groups
\cite{fb,sigl2,us2}.}\cite{us1,early}.

A generic, robust prediction of massive spin-2 exchange,
known for many decades, is that the total cross section should grow with
a power of energy, typically $\sigma_{tot} \sim s^2$. 
The property
of power law growth with a power exceeding 1 (the result of 4-Fermi
{\it spin-1} exchange) is quite hard to evade and can be traced to
dimensional analysis.  We consider large cross sections at $UHE$ 
to be characteristic of extra-dimension, low scale gravity models.
Interaction of $UHE$ neutrinos is a quite natural domain to seek the
new effects of low scale gravity models:  the very weakness of the
Standard Model neutrino coupling minimizes this background, while the
regime of of highest possible energy maximizes the effects of graviton-KK
mode exchange.   

The theory of low-scale gravity models is only partly developed, and
questions of unitarity complicate the interpretation of perturbation
theory \cite{gw}.  One can choose models of the cross section which
are further from the calculations of perturbation theory 
in the sense that they grow at a slower rate with energy than $s^2$
(The perturbative, parton level cross section rises as $\hat s^3$)
\cite{N+S,kp,usnote}, or which
operate by a separate (s-channel) mechanism \cite{d+d}.  Indeed it is
possible to restrict models of low-scale gravity to the extent
that nothing observable is predicted at the energies in question.
For example, the astrophysical bound on the scale parameter $M$ for the n=2 
case guarantees that the consequences of this model are unobservable 
\cite{N+S},\cite{kp}.

There has been some confusion on this point.  Let us compare the model
we use here, taken from our previous cross section calculations \cite{us1},
with subsequent work \cite{kp}. The latter reports the result of assuming that
a finite brane tension introduces an exponential damping of higher KK modes,
providing an alternative cutoff mechanism \cite{bando}.  Like our calculation,
when $\surd s \geq M$, the
cross section in \cite{kp} rises approximately quadratically with neutrino
energy (See Fig. 1 in \cite{kp},
where $\sigma_{\nu N}$ rises by two orders of magnitude
for every order of magnitude rise in $E_{\nu}$).  Unlike ours, the calculation
there assumes n=2 only, for which SN1987a analysis makes the
restriction $M$ $\geq$ 30-70 TeV \cite{bounds}.\footnote {Citing uncertainties 
in astrophysical 
parameters, \cite{kp} considers $M$ values as low as 6 TeV.}  If $n \geq 3$
were considered, the scale could be lowered to the 2-3 TeV range and
cross section values in agreement with ours would result.  This is clear from
the trend with mass scale in Fig. 1 in \cite{kp}.  Conversely,
we could suppress our cross section to their values by raising $M$ to values
of 6 TeV and above.  Specifically, we find that $\beta = 1$ and $M$ = 6 TeV
yields $\sigma_{\nu N}$ = 0.3 mb, compared to 0.1 mb at
$E_{\nu}$ = $10^{20}$ TeV for $M$ = 6 TeV in \cite{kp}, while the choice
$\beta$ = 2 and $M$ = 6.6 TeV or $\beta$ = 1 and $M$ = 7.3 TeV
reproduces their 0.1 mb value.  Within modest parameter variations,
the results clearly agree.  This is not a surprise, since the parton level
amplitudes and cross sections for small t are essentially identical,
and are insensitive to the value of n in the two cutoff methods \cite{bando}.
The two calculations differ only in the details of the large t cutoffs,
both of which produce $s^{2}$ behavior of the cross sections.  The
cutoff used in in \cite{kp} gives a cross section result in essential agreement
with ours at a given set of $E_{\nu}$ and $M$ values.
Their assertion to the contrary is an unfortunate consequence of
presenting the results for a lower bound $M \geq 6$, justified only for n=2,
and drawing sweeping, unjustified conclusions about the general situation.
With this technical issue clarified about the particular model we employ,
we reiterate that our goal is to 
explore the broad consequences of strongly interacting $UHE$ neutrinos.
 
The primary, general question that arises, defining our focus here,
is the nature of {\it  air shower
development}.
 High-energy leptonic interactions do not have
the same multiplicity or inelasticity as high energy hadronic
interactions.  To understand the potential relevance of neutrino
interactions, one must address not only the total cross section, but
also the way the interaction delivers energy into the air showers that
are actually observed.

In the present paper we compare simulations of air showers generated
by neutrino primaries with large cross sections to those
generated by protons in the Standard Model.  We ask whether
there is anything about existing showers which might {\it rule out}
neutrinos as primaries.\footnote{Alternatively one might ask
whether one can ``find evidence'' for neutrinos as the primaries in
some features of the showers.  We do not pursue this, because the
fluctuations of air showers and flexibility in simulation codes make
it a very hard and ambiguous way to proceed.}  
If so then the case is made that large
cross sections alone are not enough to support the case for neutrinos,
and speculative models of \emph{new} particles might be indicated.  
 
Contrary to some expectations \cite{kp}, we
find that neutrinos with large cross sections {\it can create showers
that are much like proton-initiated showers} and
that in some cases are indistinguishable from them \cite{sm} 
\cite{ddm}. Two features of the low-scale gravity contribution to the
neutrino-nucleon cross section come into play: first, the cross section is
large enough to initiate air showers at high enough altitudes; second, its
rapid $s^2$ dependence suppresses new effects among secondary products,
which carry at most a few percent of the primary energy.  Our  methodology 
can evidently be extended to other models for hadronic size UHE neutrino
cross sections provided the cross sections grow rapidly (as in spin-2 
exchange).  Other speculative primaries should be considered on a case-by-
case basis.  One cannot take the existence of one model that
produces well simulated
showers above $10^{20}$ eV to be conclusive evidence for a given
hypothesis for the identity of the primary, neutrino or otherwise.  The
question of the mysterious primary, then, needs to be framed in view
of everything that can be observed:  cross sections, shower
characteristics, and angular distributions and correlations, which
may be informative about the charge of the primary. 

%

\section{Air Showers with Neutrino Primaries}
The neutrino proton cross section is significantly modified at ultra
high energies due to graviton exchange within low scale gravity models
\cite{add}.
Within these models the Feynman rules can be found in \cite{HaLyZh}.
These rules, derived for the case of a common compactification scale
for all compact dimensions,
are applicable only at energies smaller than the fundamental
scale of quantum gravity $M$. We are not primarily concerned here with
variations on the theory, which can raise and lower scales somewhat,
\footnote{Experimental bounds are generally restricted to the case
where a common radius is assumed.  This restriction is convenient,
but not necessary, as remarked for example in \cite{a+b-pascos}.} but
we do assume that any new physics has a scale of about $1$ TeV.
\footnote{We are not considering models where gauge and matter fields
propagate in the extra dimensions.  Bounds that apply to such models
and to the n=2, common radius model of low scale gravity are reviewed in
\cite{nath}, for example.}
Experimental limits on the effective scale in the theory 
depends on the number of extra dimensions in the Universe.
If the number of extra dimensions is larger than 2,
and a common compactification
scale is assumed, then $M$ is constrained to be larger than about 4 TeV,
1 TeV and 0.5 TeV for n = 3, 4 and 6. \cite{bounds}. Given uncertainties
in the estimates, these are all acceptable for our purposes.
The energies involved
in the ultra high energy cosmic ray events are much larger than these scales.
In order to extend our calculations beyond the scale $M$ some
modelling is required, since the calculational procedure beyond this
scale within quantum gravity is unknown.
 
Our procedure is to make calculations with several different models.
At the parton level above $\surd s \simeq M$, the cross sections rise
with $\hat s$ either as $\hat s^n$, with $n=1,2$ or as $(\log \hat s)^2$.
A natural feature of the perturbative $\hat s^3$ growth of the spin-2 exchange 
below the scale $M$ is that the effects of new
physics lie \emph{well below} the sensitivity of accelerator experiments
below this scale.
The new cross section effects rise to become comparable to the Standard Model
alone at about $\surd s \simeq M$, as expected. The total cross section
then rises quickly above the Standard Model above $\surd s \simeq M$.
Depending on the choice of $M$
and the model used we found that the cross sections range from 1 mb
to several hundred mb at energies of the order of $10^{20}$ eV \cite{us1}.  

We developed a program that can generate air showers with a non
standard neutrino primary using the AIRES and PYTHIA simulators.
The steps in this Monte Carlo simulation are as follows: (1)  
The neutrino collides with an air nucleus at an altitude which is
determined by the scattering cross section. The neutrino proton
cross section is calculated by using the methods explained in
Ref. \cite{us1}. The neutrino-nuclei cross section can then
be computed using the standard Glauber formalism \cite{glaub}. (2) The 
neutrino typically loses less than $10\%$ of its energy in any of these
collisions, a point which has earlier been emphasized by \cite{kp}.
The neutrino typically undergoes collisions with several protons 
in the first nucleus as well as the subsequent ones it 
hits. The number of hits on target nucleus of weight A is given by
\cite{Gaisser}
$$n_{\rm hits} ={ A\sigma_{\nu p}\over \sigma_{\nu A}}$$ In step (3),
for each of the hits we determine the outgoing particles by using PYTHIA. 
(4) We stack all the final state particles produced by the PYTHIA simulation 
except those which originate from the
decays of $\pi^0$ and $K^0$. We inject $\pi^0$ and $K^0$ directly since these
can be processed by AIRES.  These are stacked into AIRES at each point
that the neutrino proton collision occurs. This sequence is then reproduced
probabilistically over the course of the shower.

Let us emphasize again that our primary interest here is the development of
showers initiated by a large cross section, low inelasticity, neutral-
current-like primary interaction.  Models where only the neutrino-hadron
interaction feels the new physics can readily be treated with our analysis.
In the particular class of models that
we consider, the hadronic interactions of the secondaries are, in principle,
also affected.  In practice, these effects are not important.  The energy
transfer to the hadron system is less than $10\%$ of $E_{\nu}$ per collision.
The multiplicity at $10^{20}$ is of order 100.  We find that the nominal
value of $0.01\times E_{\nu}$ per secondary can fluctuate up to as much
as $0.05\times E_{\nu}$ for one or two secondaries.  This still leaves
the highest energy secondaries with less than $10^{19} eV$ ``lab energy''.
Even the largest cross sections we consider are $\leq$ 1 mb in this energy
range, much smaller than the expected values of order 100 mb for the
SM hadronic cross sections.  It is clear that the showers will not be
significantly affected by changes of $1 \%$ in one or two particles in the
shower. For this reason we do not include the KK graviton excitation
corrections to the secondary hadron interactions.

\begin{figure}
\subf{
\bmi[t]{6in}
\includegraphics[scale=1.00]{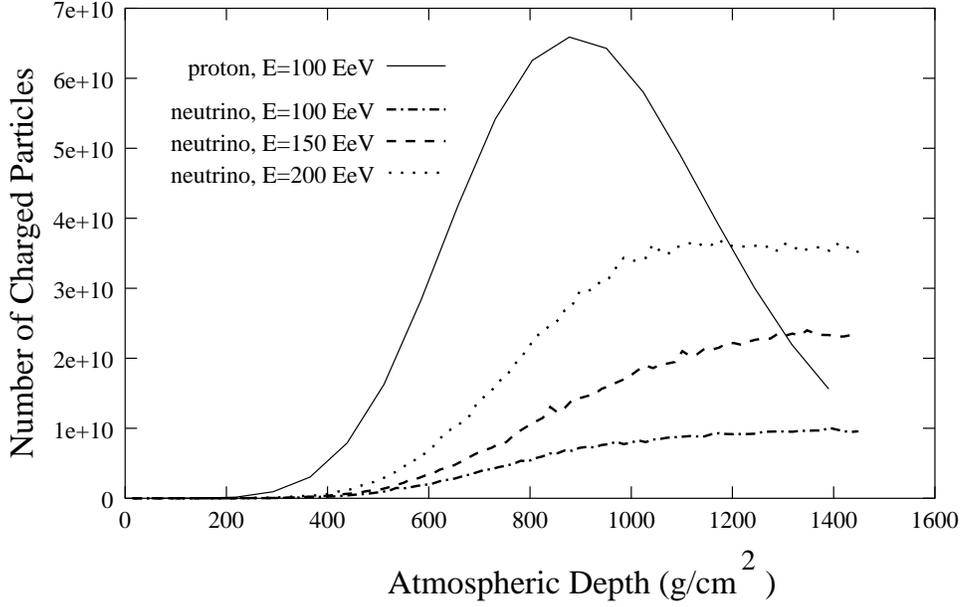}
\emi
}
\caption{The longitudinal shower profiles averaged over 50 showers 
for the case when the neutrino proton cross section $\sigma_{\nu p}$ 
is less than about
20 mb at the ultra high energies of the order of 100 EeV. The longitudinal
profile for neutrino induced shower profiles 
with primary energy E=100 EeV (dash dot curve), 150 EeV 
(dashed curve) 
and 200 EeV (dotted curve) are compared with a proton induced shower
with primary energy of 100 EeV (solid curve). The $\sigma_{\nu p}$ for this
case is obtained by using the linear rise model, $\hat \sigma\propto \hat s$,
where $\hat\sigma$ and $\hat s$ are the parton level cross section and
center of mass energy respectively. 
The $\sigma_{\nu p}$ 
values are 9.4 mb, 15.3 mb and 21.5 mb for primary energy E=100, 150
and 200 EeV respectively.  
}
\label{linear}
\end{figure}

In our simulations we study the quantum gravity parameter space such that
the neutrino proton cross section ranges from about 10 mb to several
hundred millibarns.  This is a reasonable range, suggested by bounds based
on experiments.  There are two notable regimes:

*If we lower the cross section below about 20 mb 
we find that the air showers generated are very different from
those initiated by a proton.  For instance, the showers are stretched out
by $50\%$ or more, with the location of shower maximum delayed by a similar
amount\footnote{These smaller cross section values are interesting from the
point of view of horizontal shower searches \cite{tos}.}. This behavior is 
illustrated in Fig. \ref{linear}, 
where the shower profile averaged over 50 showers
for a proton primary with energy $100$ EeV is compared to the profiles
for a neutrino primary with linearly rising cross section and energies of
$100$, $150$ and $200$ EeV. The neutrino proton cross sections in this case
are 9.4 mb, 15.3 mb and 21.5 mb for neutrino energy E=100, 150
and 200 EeV respectively. 
 Though interesting in their own right, these
cases will not be readily confused with the observed highest energy showers.

*If the cross sections are much larger than about 20 mb, then a variety
of things can occur.  When the neutrino - proton cross section is about 
the same as the proton - proton cross section, we find that the showers 
generated by neutrinos \emph{may or may not} differ in detail from those
generated by protons.  There is always a region of parameter space where
the difference in showers is too small to detect.         
We therefore  concentrate this study on the larger cross section
values attainable with the characteristic $s^2$ ``Regge'' rise\cite{us1}.

Compared to a proton, the neutrino loses a small amount of energy
per collision. For this reason a neutrino undergoes collisions with
many air nuclei. The interaction basically occurs along a line rather
than at a single point\footnote{This effect, which explains the observed
spread in arrival times of particles far from the core \cite{nw}, is
exaggerated in neutrino induced compared to proton induced showers.}.
A proton shower will arise primarily from the proton's collision with a
single air nucleus, since the energy loss per collision is large 
and the secondary showers generated by collision of the remnants of the
incident proton with other air nuclei will be relatively weak. 
The shower-to-shower fluctuations can still be large, however, and detailed
study is required.

\section{Results and Discussion}

\begin{figure}
\subf{
\bmi[t]{6in}
\includegraphics[scale=1.00]{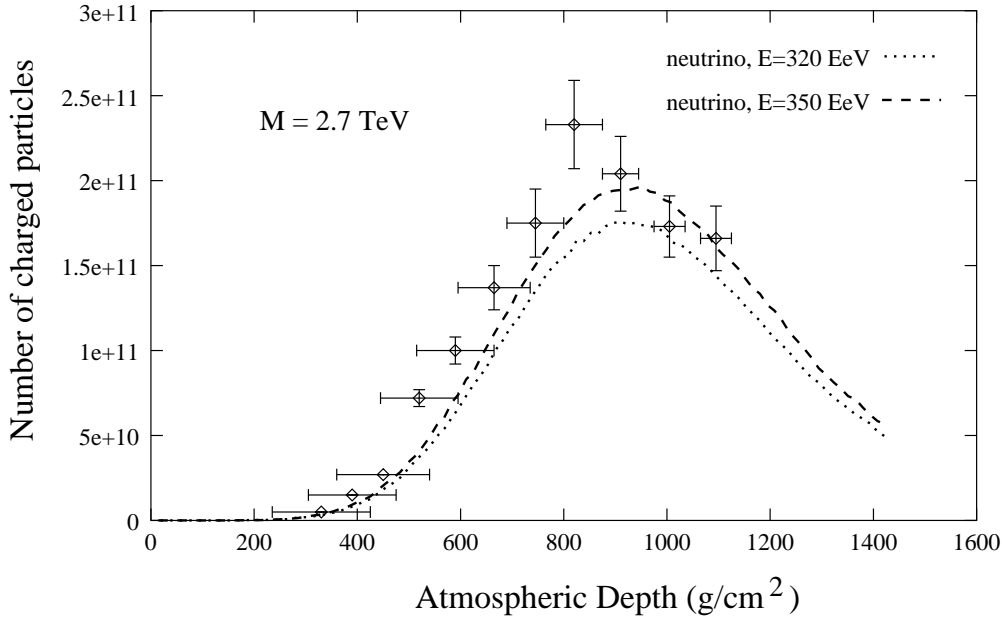}
\emi
}
\caption{
The longitudinal profile of showers generated by neutrino primaries
compared to the Fly'e Eye data. The dotted and dashed 
curves show mean over 50 showers
with neutrino primary energy $E=320$ EeV and $350$ EeV respectively.
The quantum gravity scale $M=2.7$ TeV is used in these simulations.
}
\label{FE320_1}
\end{figure}

We compare the structure of neutrino induced showers to proton induced
showers using the ``Regge model'' cross section mentioned above \cite{us1}.
We also show in Fig. \ref{linear} 
a few selected cases of showers induced by the linear-in-s
cross section case to illustrate that lower cross section, deeper and
more extended events are rather distinct.  
The choice $\beta = 1$ is sufficient for our study, and all of the plots
from Fig. \ref{FE320_1} onwards
are made with this value\footnote{The momentum transfer has a cutoff
$1/(M^2-\beta t)$ in this model.}.  The injection energy of the neutrino
and the value of $M$ are then chosen to create a shower to be compared
with data or with a simulated proton shower of prescribed energy.
                    
\begin{figure}
\subf{
\bmi[t]{6in}
\includegraphics[scale=1.00]{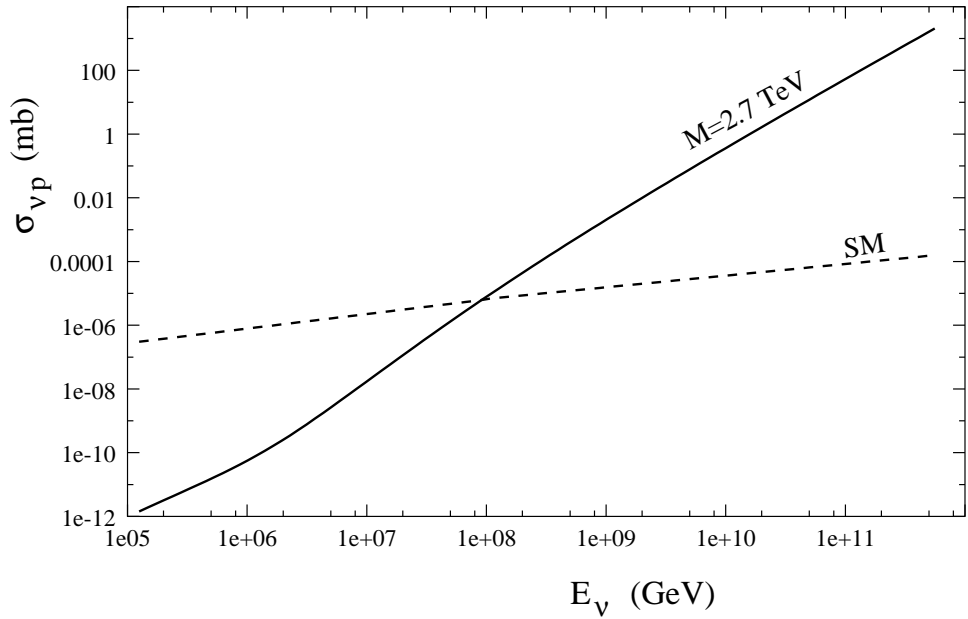}
\emi
}
\caption{
The neutrino-proton cross section, $\sigma_{\nu p}$, in large extra
dimension models assuming $s^2$ growth of parton level cross section
with the quantum gravity scale $M=2.7$ TeV (solid curve) compared to
the standard model result (dashed curve). 
}
\label{cross_section}
\end{figure}

\begin{figure}
\subf{
\bmi[t]{6in}
\includegraphics[scale=1.00]{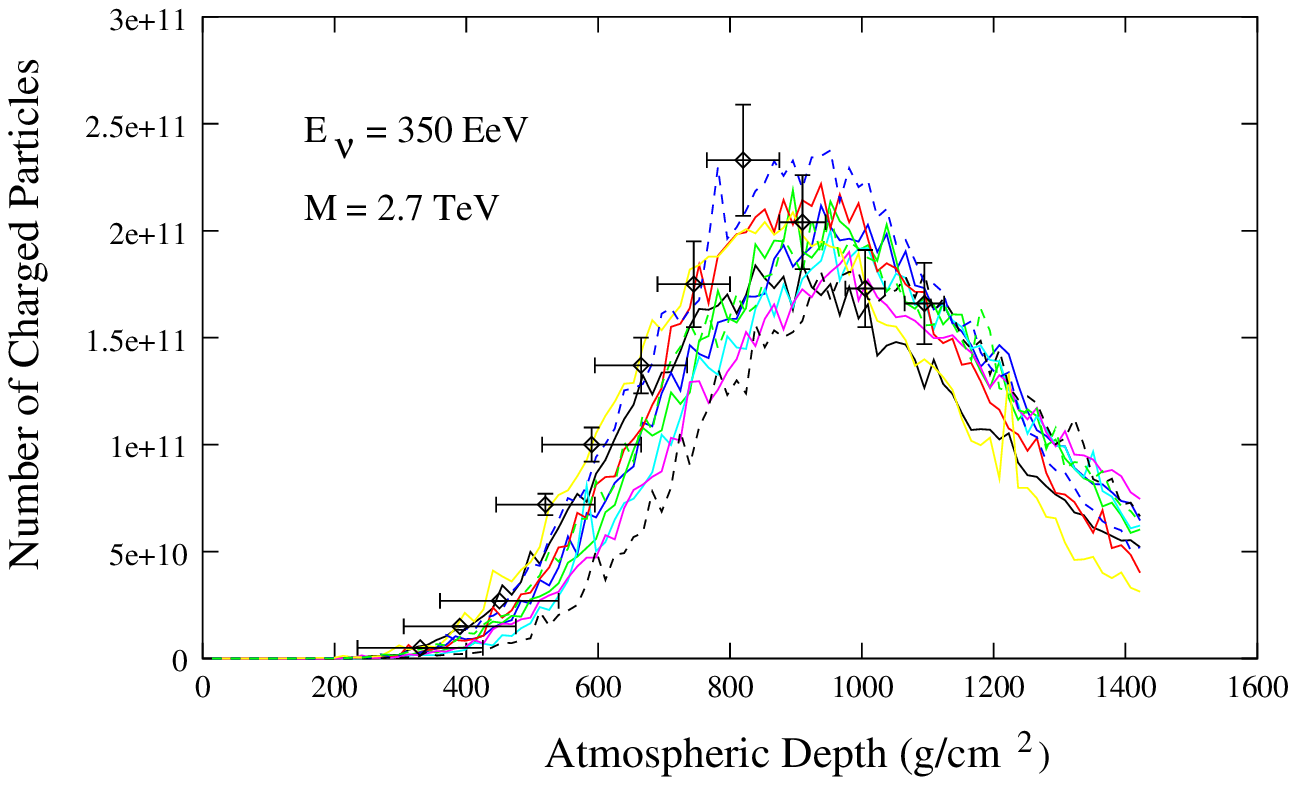}
\emi
}
\caption{
The longitudinal profile of 10 showers generated by neutrino primaries
compared to the Fly'e Eye data. The neutrino primary has energy $E=350$ EeV
and the quantum gravity scale $M=2.7$ TeV is used in these simulation. 
}
\label{scatterFE320}
\end{figure}

Of all the ultra high energy cosmic ray detectors, only the
Fly's Eye and its offspring HiRes track longitudinal development of the
showers.  To start our comparison of neutrino and proton initiated showers,
we show in Fig. \ref{FE320_1} the profile, 
or number of charged particles vs. depth, of the
highest energy cosmic ray event ever observed. The data points  
are the best reconstruction of the event, as
analysed and presented by the Fly's Eye group \cite{320}. The energy is
quoted as $320$ EeV, and we superpose the profiles averaged over 50 simulated
neutrino-induced showers for the case $M = 2.7$ TeV and energies $E_{\nu}$ =
$320$ EeV and $350$ EeV.  
The corresponding neutrino-proton cross section 
with the quantum gravity scale $M=2.7$ TeV
is shown in Fig. \ref{cross_section}. 
As noted by Bird et al. \cite{320}, the best fit 
value of the depth of the shower maximum is consistent with the primary being
a proton, midsize nucleus, heavy nucleus or even a gamma ray.  Could it have 
been a neutrino?  In Fig. \ref{scatterFE320} 
we show the same data and the profiles of 10
simulated individual neutrino showers with $M = 2.7$ TeV and $E_{\nu}$ =  
$350$ EeV.  The shower-to-shower fluctuations are vividly illustrated, with
the envelope doing a good job of capturing the event.  From this example,
we  would say a neutrino primary with a large cross section but a small
energy transfer per collision like a neutral current interaction and an 
energy of about 350 EeV is not ruled out.  
 
\begin{figure}
\subf{
\bmi[t]{4in}
\includegraphics[scale=0.65]{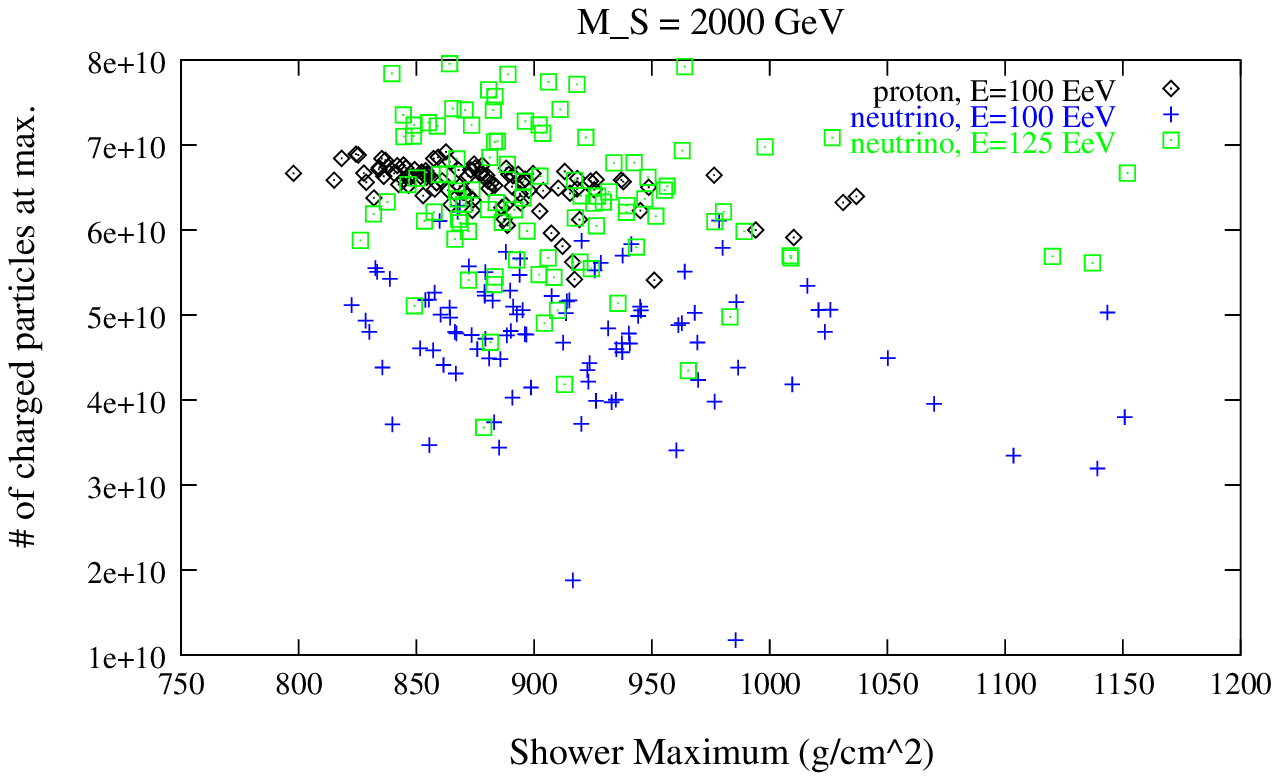}
\emi
}
\subf{
\bmi[t]{4in}
\includegraphics[scale=.65]{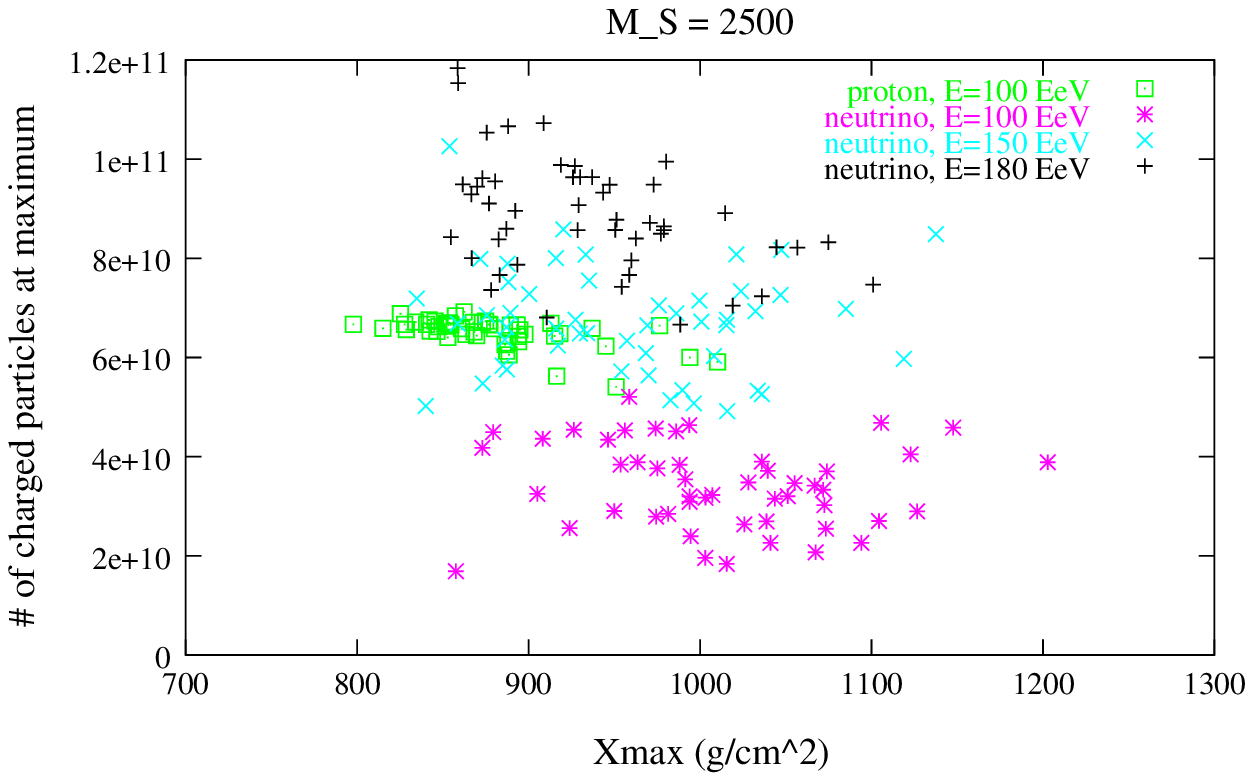}
\emi
}
\subf{
\bmi[t]{4in}
\includegraphics[scale=.65]{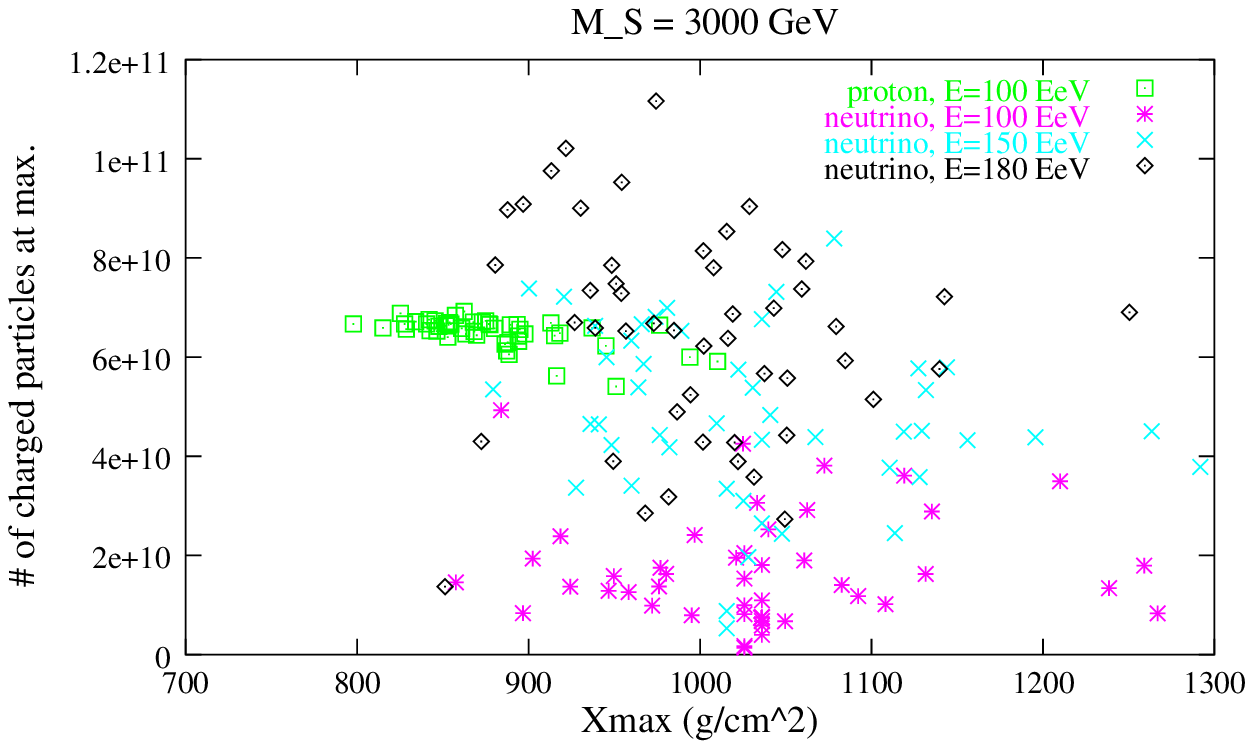}
\emi
}
\caption{The number of charged particles at shower maximum ($N_{max}$)
vs. the atmospheric depth of the maximum ($X_{max}$) for several different
choices of the quantum gravity scale $M$ and the incident neutrino energy.
The result for proton ($E=100$ EeV) induced showers within 
the standard model are shown
for comparison.  Fifty showers are shown in each case.
} 
\label{Nmax}
\end{figure}

To develop the points of comparison suggested by Figs. \ref{FE320_1} and 
\ref{scatterFE320}, 
in Fig. \ref{Nmax}   
we show the scatter plot for 50 showers of
the shower maximum ($X_{max}$) versus the
number of particles at maximum ($N_{max}$) for both the proton and neutrino
primaries for several assumed values of the incident neutrino primary
energy and the scale $M$.  The proton shower energy is fixed at 
$E=10^{20}eV$, while the initial neutrino energy is varied for each of the
values M=2, 2.5 and 3 TeV chosen.
We find that the neutrino showers show more scatter
than do the proton showers. The scatter decreases as the 
value of $M$ decreases i.e. as the $\sigma_{\nu p}$
{\it increases}.  Although the mean values for these two 
types of showers differ for most of the parameter space there exist
many neutrino generated individual showers which are very 
similar to the proton shower with $E=100$ EeV.
For $M=2.0$ TeV we find that there is somewhat more scatter but the
same average $N_{max}$ and $X_{max}$ values for neutrino showers with 
primary energy 
$125$ EeV compared to proton showers with primary energy of $100$ EeV.
Almost the same statement applies to the comparison of neutrino showers where
$M=2.5$ TeV and $E=150$ EeV are compared to the proton sample, though the
$X_{max}$ value is a bit higher in this case.

\begin{figure}
\subf{
\bmi[t]{4in}
\includegraphics[scale=0.65]{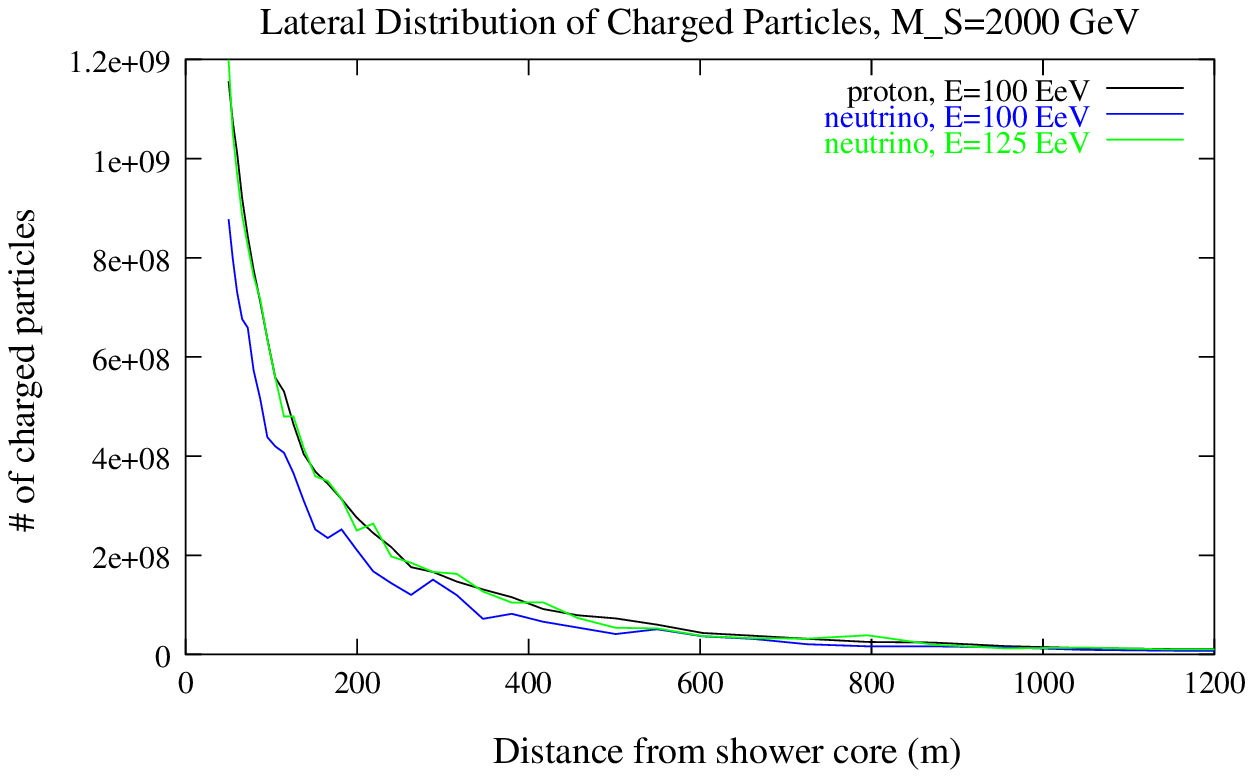}
\emi
}
\subf{
\bmi[t]{4in}
\includegraphics[scale=.65]{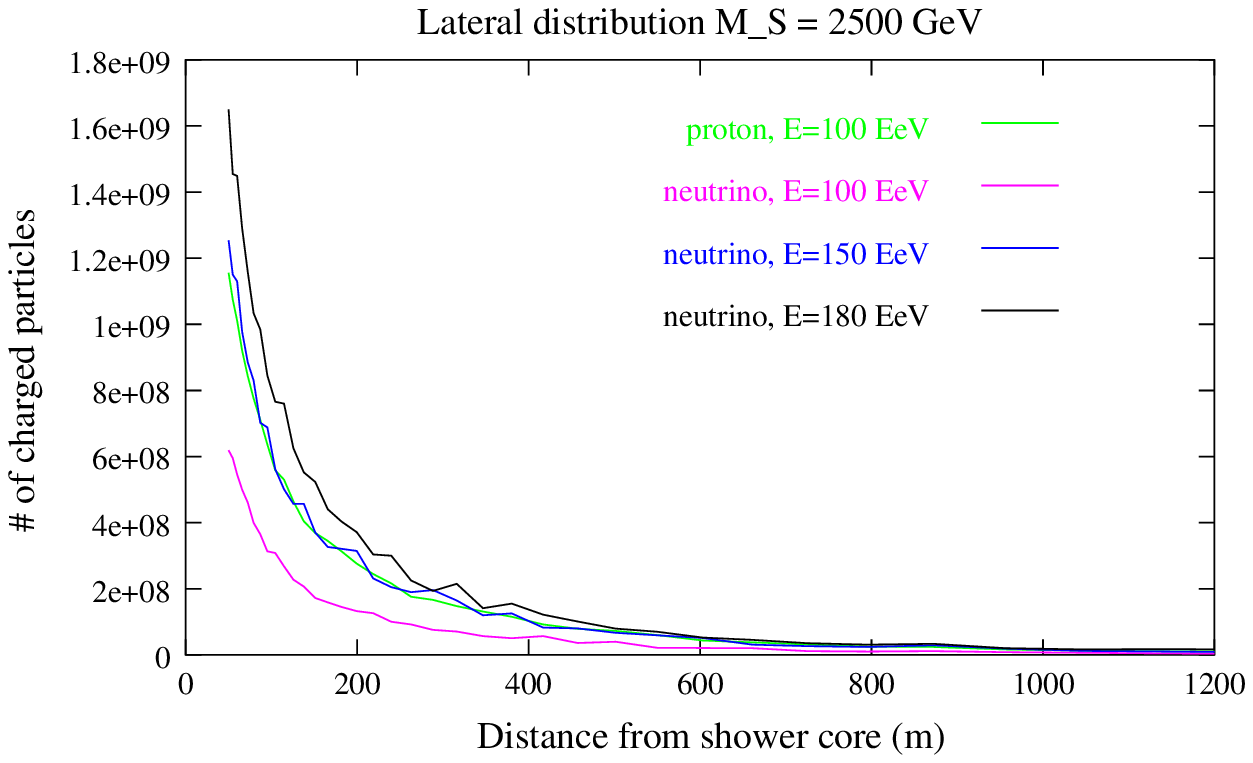}
\emi
}
\subf{
\bmi[t]{4in}
\includegraphics[scale=.65]{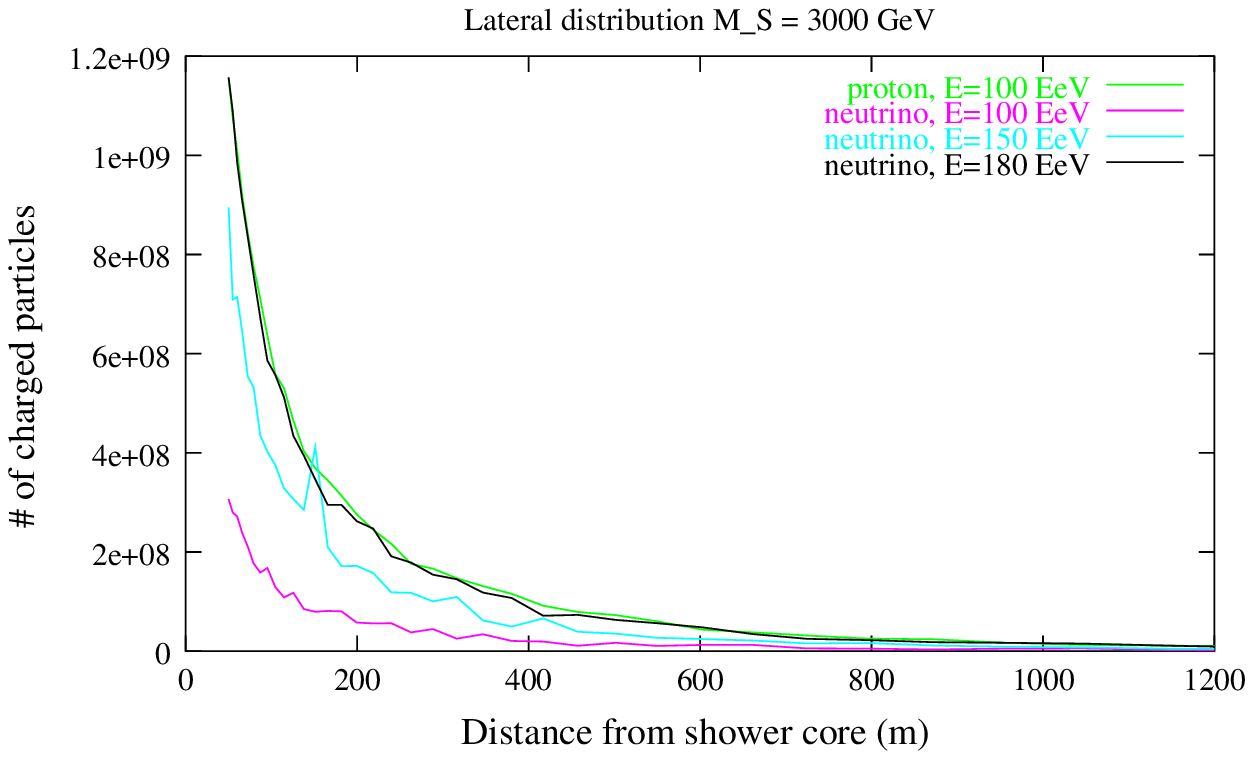}
\emi
}
\caption{The lateral distributions of charged particles 
at ground level for several different
choices of the quantum gravity scale $M$ and the incident neutrino energy.
The result for proton ($E=100$ EeV) induced showers 
within the standard model are shown
for comparison.  Each curve is an average over 50 showers.
} 
\label{lateral}
\end{figure}

As mentioned above, the Fly's Eye and HiRes, using the air flourescence
technique, are the only experiments that directly reconstruct the profiles.
The others, the Volcano Ranch array, the Haverah Park array, the Sydney
University array (SUGAR), the Yakutsk array and the Akeno Giant Air-Shower
Array (AGASA) all employed particle detection schemes to
observe the lateral pattern of the shower particles at
ground level. For example, AGASA deduces the energy of the primary
by measuring the density of charged particles at 600 m from the shower
core.  Averaging over 50 showers, we compare this fundamentally important,
lateral distribution of charged particles at
ground level for the two types of showers in Fig. \ref{lateral}. 
We find that the two distributions are \emph{very similar} for a number
of neutrino shower $M$ and $E_{\nu}$ choices and $E_{p}$ = $100$ EeV.  

\begin{figure}
\subf{
\bmi[t]{4in}
\includegraphics[scale=0.65]{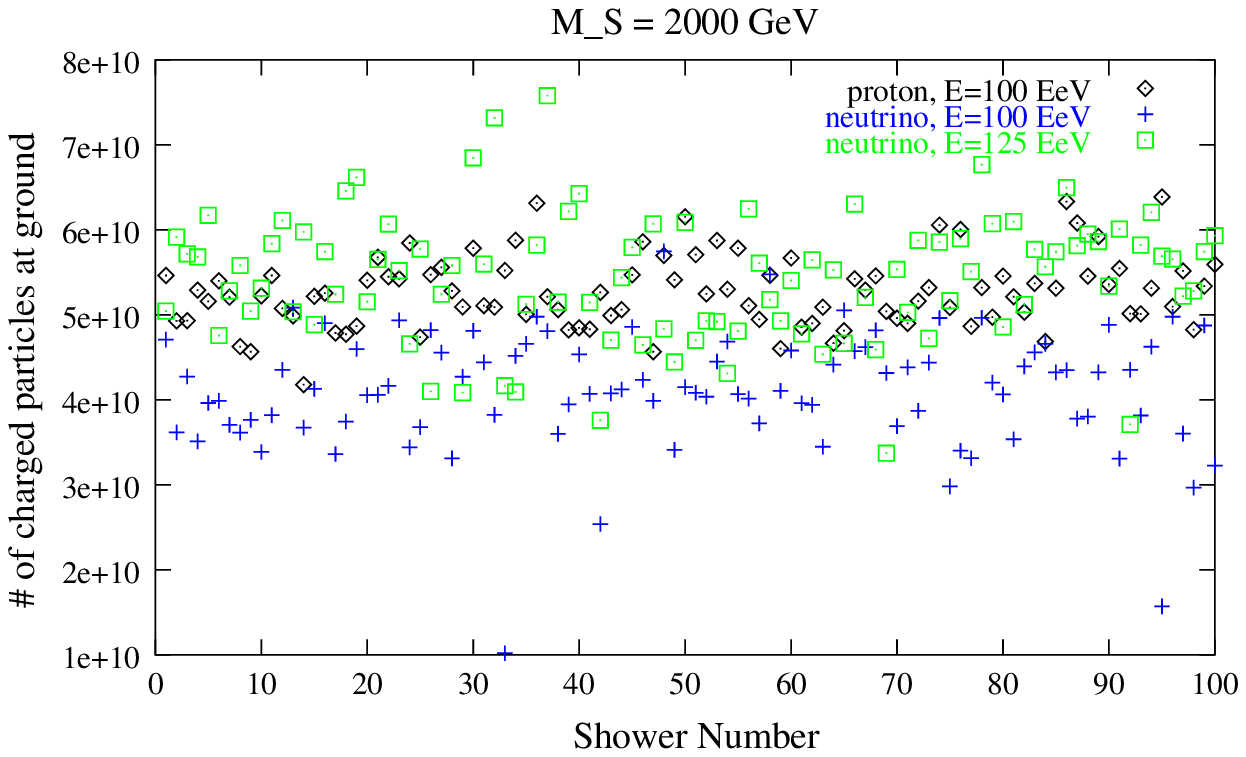}
\emi
}
\subf{
\bmi[t]{4in}
\includegraphics[scale=.65]{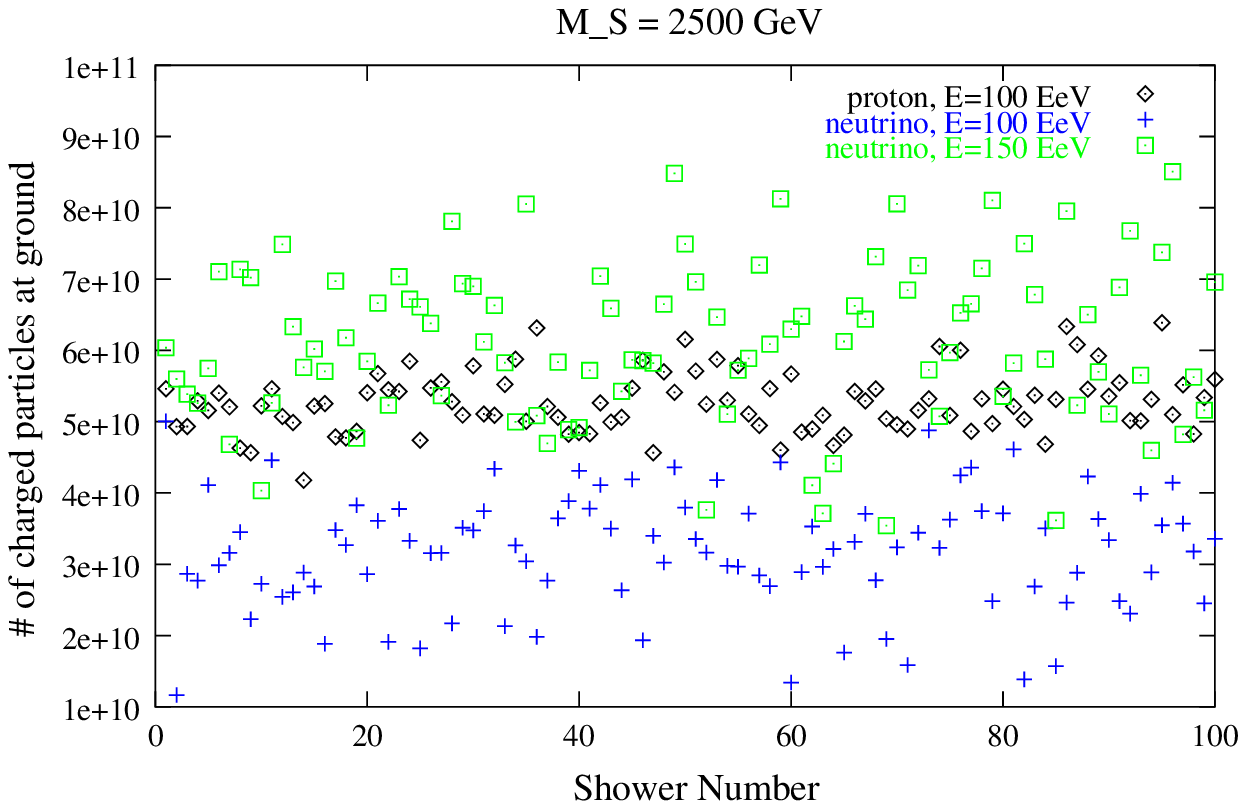}
\emi
}
\subf{
\bmi[t]{4in}
\includegraphics[scale=.65]{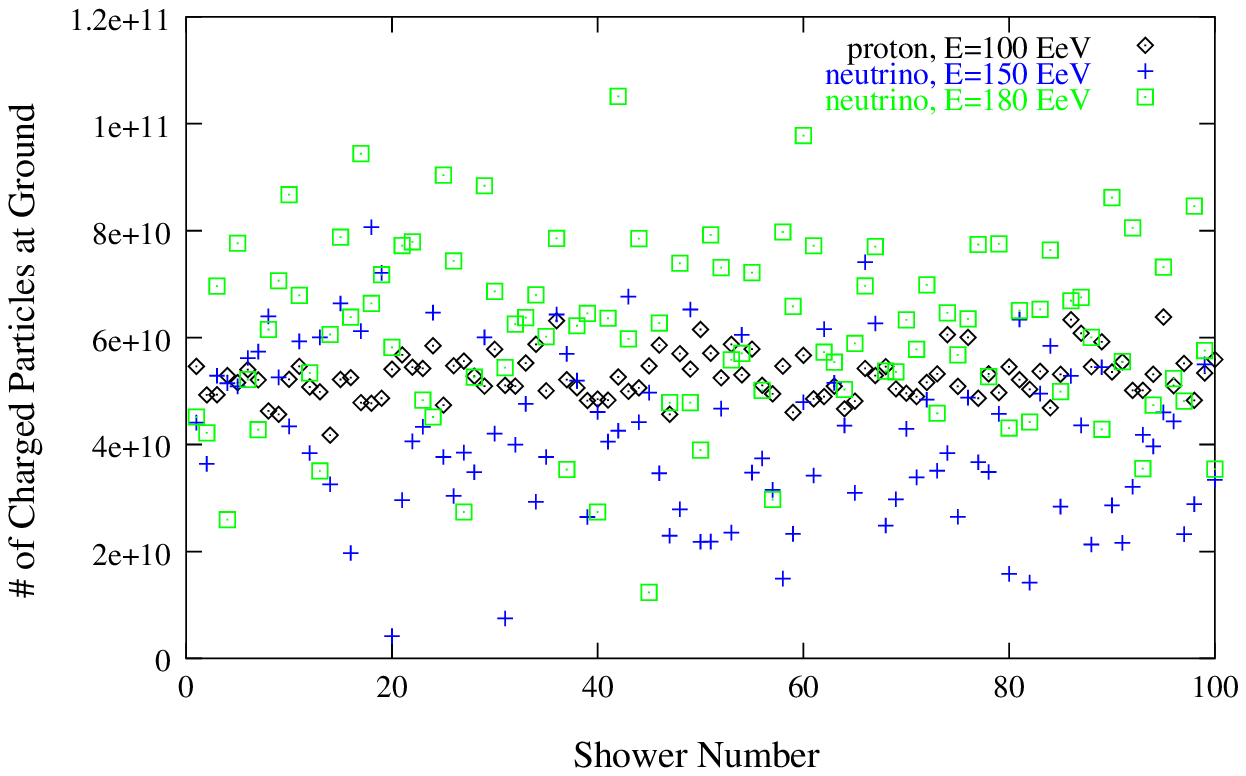}
\emi
}
\caption{The number of charged particles 
at ground level for several different
choices of the quantum gravity scale $M$ and the incident neutrino energy.
The result for proton ($E=100$ EeV) induced showers 
within the standard model are shown
for comparison.
} 
\label{grdpcles}
\end{figure}

We next compare the total number of charged particles at ground level
for the proton and neutrino generated air showers. The results are
shown in the scatter plot, 
Fig. \ref{grdpcles}. 
We find that the total number of particles at ground
level is in general smaller for a  neutrino in comparison to a proton
primary if both the incident particles have the same energy. This
difference disappears if the $\sigma_{\nu p}$ is very large (equivalently
$M$ is small).  For a given value of $M$, one can raise the neutrino
energy and see the number of particles at ground level increase toward
the value of the proton shower.  For example,  
for $M=2.0$ TeV we find that the average number of charged
particles at ground level for neutrino showers is the same as for proton
showers if the incident neutrino has a $25\%$ higher energy. 
For larger values of $M$, of the order of $3.0$ TeV, we find
that the number of charged particles at ground level is again the
same as that of a proton induced shower with primary energy 100 EeV, 
if the neutrino primary has energy roughly equal to 180 EeV. Based on
this diagnostic alone, these results suggest that 
AGASA might interpret a shower generated by a neutrino
primary to be that generated by a proton of a somewhat smaller energy.
 
\begin{figure}
\subf{
\bmi[t]{4in}
\includegraphics[scale=0.65]{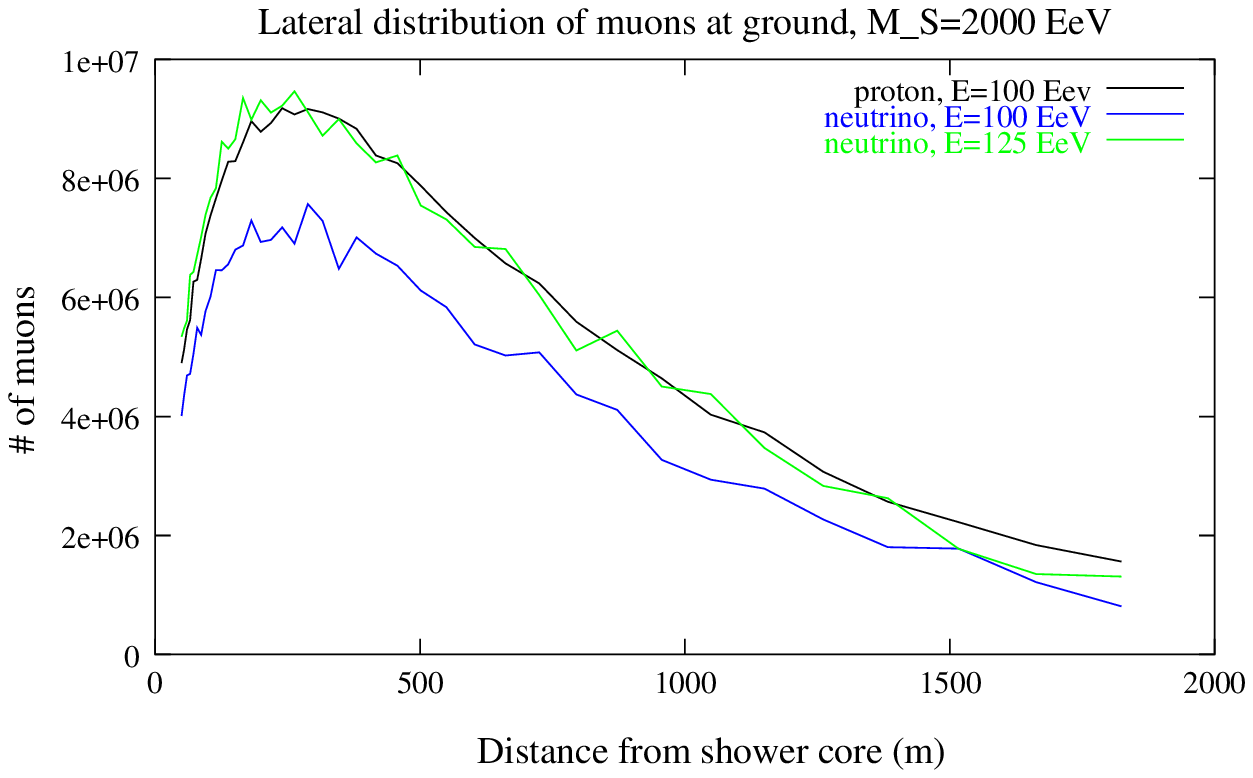}
\emi
}
\subf{
\bmi[t]{4in}
\includegraphics[scale=.65]{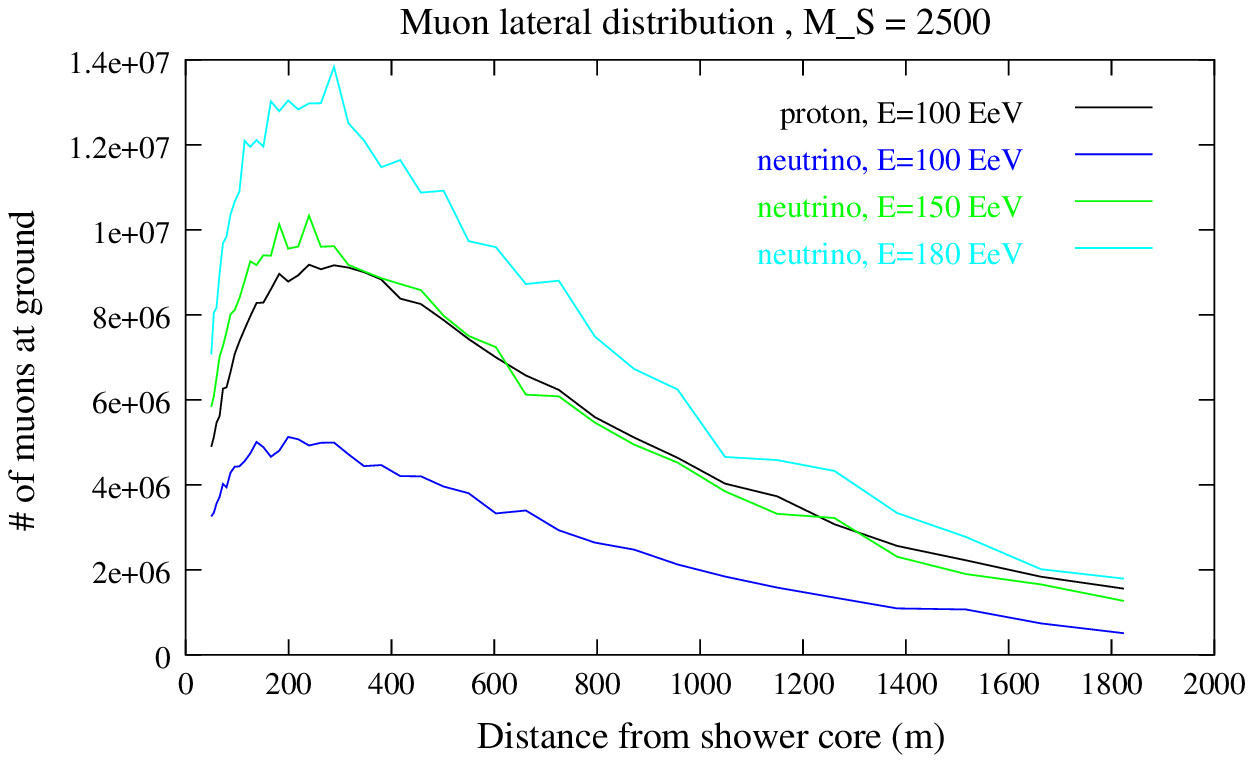}
\emi
}
\subf{
\bmi[t]{4in}
\includegraphics[scale=.65]{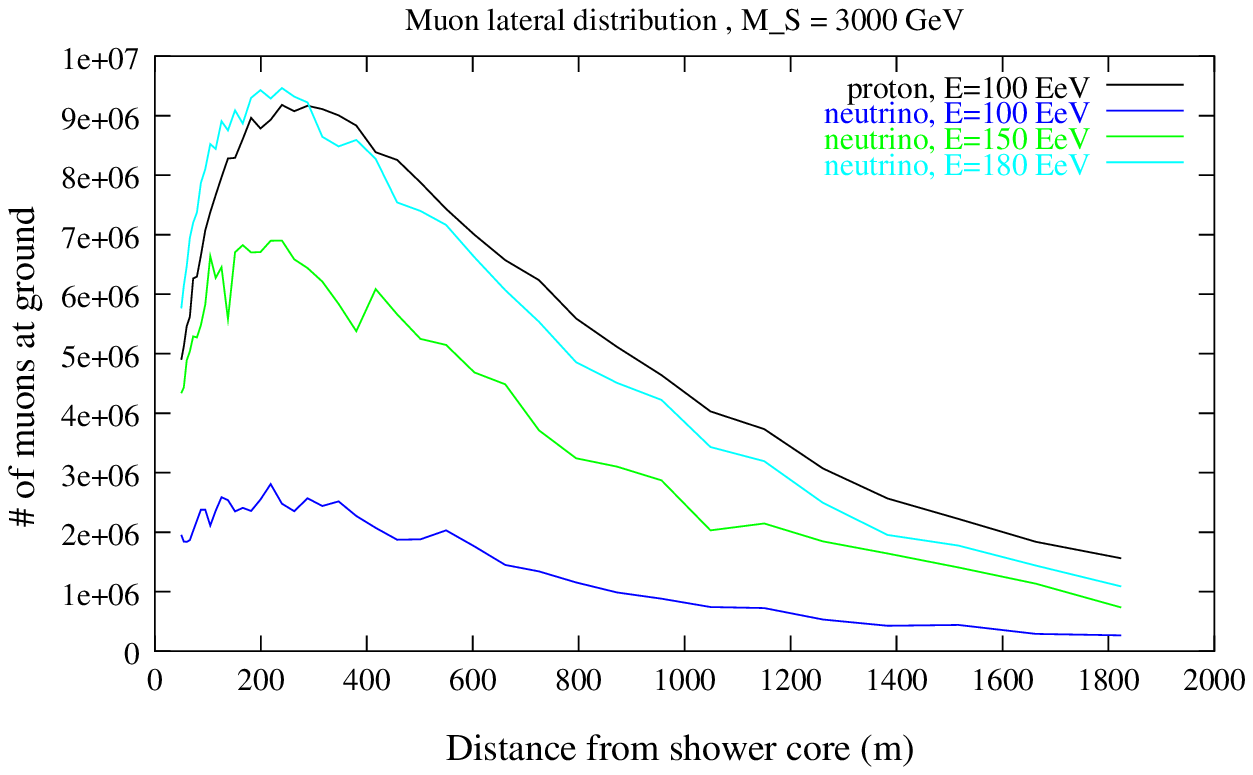}
\emi
}
\caption{The lateral distribution of muons at ground level 
for several different
choices of the quantum gravity scale $M$ and the incident neutrino energy.
The result for proton ($E=100$ EeV) induced showers 
within the standard model are shown
for comparison.  The curves show averages over 50 showers.
} 
\label{muon}
\end{figure}

The identity  of the primary particle (p, Nucleus or $\gamma$)
is deduced by AGASA on the basis
of the muon content of the shower. The lateral dependence of the 
number of muons observed at ground level, again averaged over 50 showers,
is shown in Fig. \ref{muon}. 
We find that the muon distribution for the two types of primaries
is approximately the same for a number of $E_{\nu}$ and $M$ 
combinations. This diagnostic is evidently not a sensitive tool for
discriminating between the proton-induced and neutrino-induced showers.

\begin{figure}
\subf{
\bmi[t]{4in}
\includegraphics[scale=0.65]{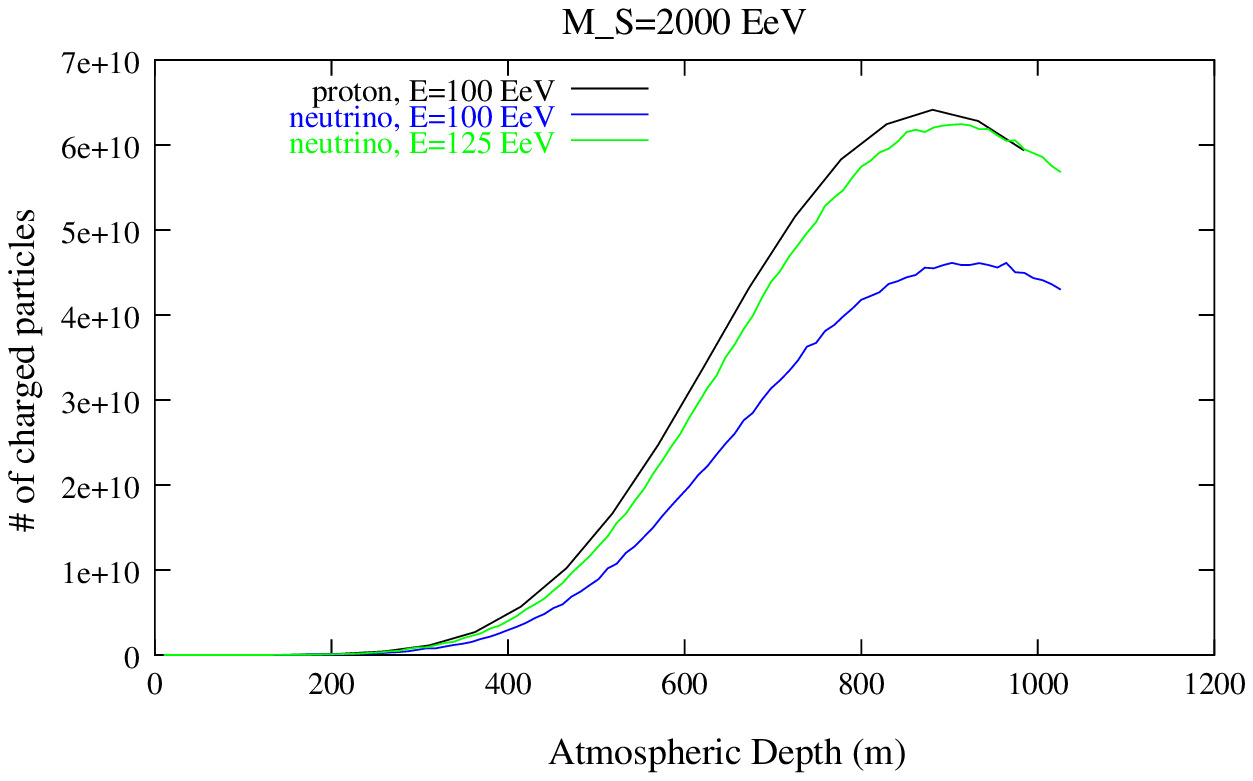}
\emi
}
\subf{
\bmi[t]{4in}
\includegraphics[scale=.65]{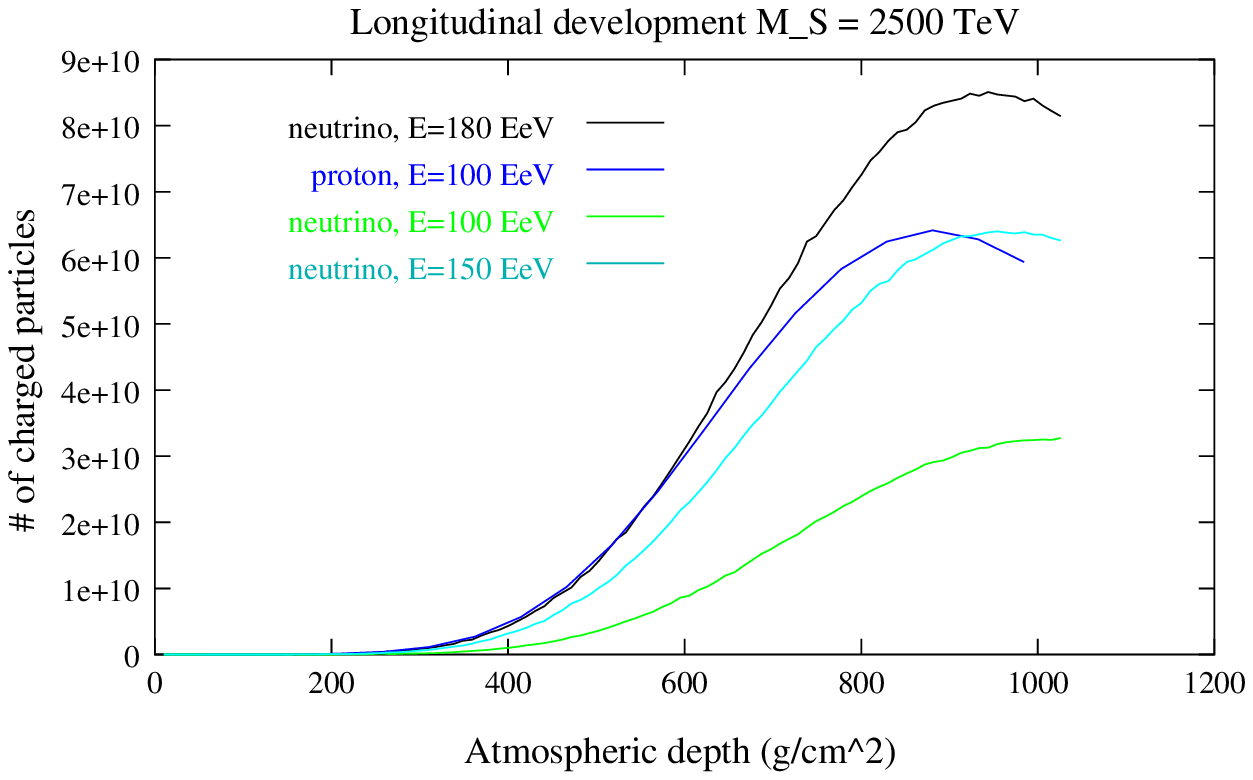}
\emi
}
\subf{
\bmi[t]{4in}
\includegraphics[scale=.65]{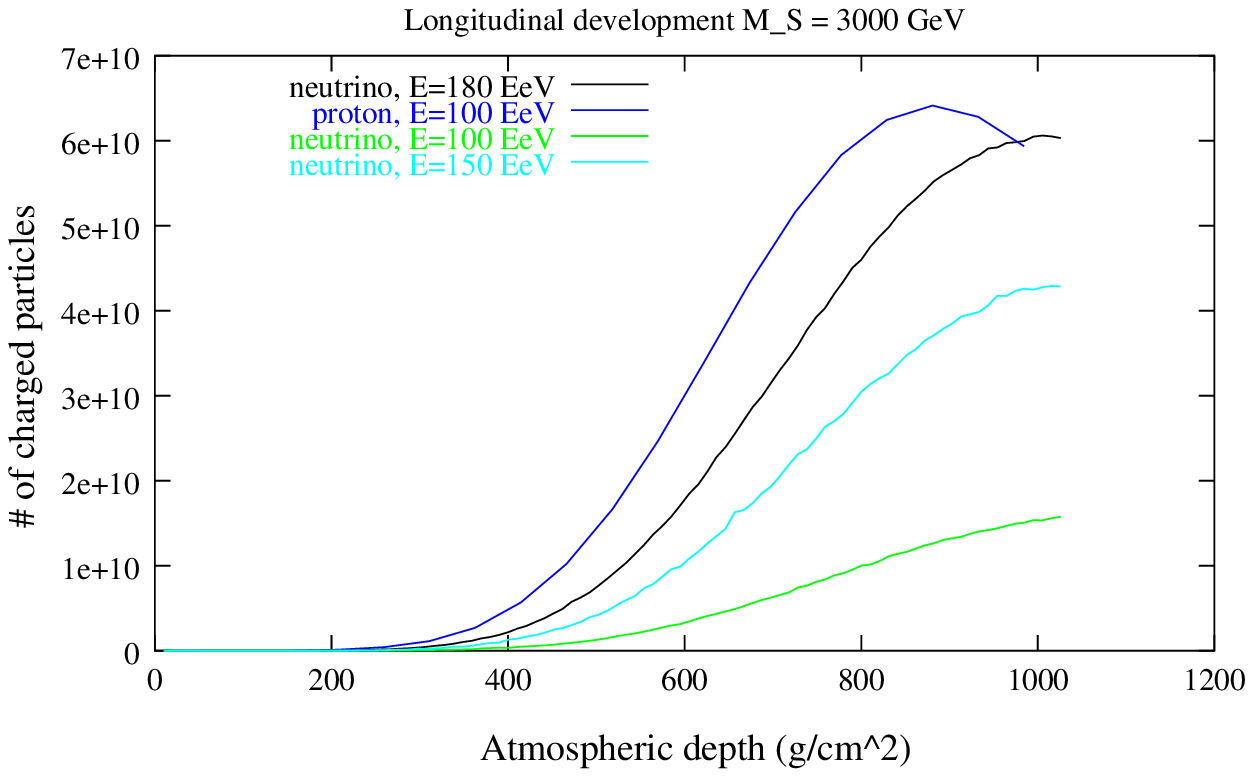}
\emi
}
\caption{The longitudinal distribution of charged particles 
for several different
choices of the quantum gravity scale $M$ and the incident neutrino energy.
The result for proton ($E=100$ EeV) induced showers 
within the standard model are shown
for comparison.  Curves show averages over 50 showers.
} 
\label{long}
\end{figure}

Coming back to the longitudinal development of the charged particles,
we show a variety of cases, averaged over 50 showers, in Fig. \ref{long}.  
In this case we find significant difference
between the showers generated by neutrino and proton primaries.
The neutrino showers in general show maximum closer to the 
ground level. This difference, however, essentially disappears
if the scale $M$ is roughly 2 TeV.  However, as one can see
in these plots and in the scatter plots of $N_{max}$ vs. $X_{max}$,
Fig. \ref{Nmax}, if $M$ is a bit above $2.0$ TeV at 2.5 or 3.0 TeV, 
then one cannot  
bring the average position of shower maximum, $X_{max}$ into line with
the average profile of proton showers of a given energy.\footnote{This
insensitivity of the depth of shower maximum just reflects the weak $ln(E)$
dependence of the maximum \cite{rossi}.}
Since shower-to-
shower fluctuations are large, as seen in Figs. \ref{FE320_1} and 
\ref{scatterFE320}, it requires

a significant sample such as the 50 - 100 shown in our study, to get
good discrimination. 

\section{Discussion and Conclusions}

The conjecture that UHE, super-GZK cosmic ray showers are caused by neutrinos
with large UHE cross sections is almost as old as the field itself
\cite{early}.  Revisiting this idea in a new theoretical framework \cite{add},
we proposed models that achieve interestingly large cross sections \cite{us1}.
We speculated that the GZK ``barrier'' could be then broken by neutrinos.
The next question to answer is whether the shower events predicted look
like events observed. Or are the characteristics so different from observed
showers, which are generally compared to those of the simulations of
proton, nucleus and gamma initiated showers, that the neutrino can be 
eliminated as a candidate for the super GZK showers?
{\it Our conclusion based on this study is no, they cannot be eliminated
as candidates}.  For a range of values of
the fundamental scale $M$ in the neighborhood of $2 -3$ TeV, there are 
neutrino energies 25\% - 75\% above that of the comparison proton model where
the simulations match quite closely.  As noted in the introduction, the same 
analysis applies to a variety of cases - different interactions and different
identities of primary particle.  Given that the same cross section input
is not unique to the low scale gravity inspiration used here, this result is
of quite general use. 

Whether neutrinos follow the standard model extrapolations \cite{q+r,Frich96},
are enhanced ``modestly'' by 3-5 orders of magnitude
or ``extravagantly'' by more than 5 orders of magnitude, the search
for UHE neutrino induced events at present and new facilities
\cite{auger,icecube}
will be an exciting one.
  
{\bf Note Added}: As we were completing this paper, a closely related
work appeared \cite{neu}.  The cross sections considered in their work
are well below 10 mb, while we concentrate primarily on cross sections
above this value.  Where cross sections are roughly the same,
results and conclusions qualitatively agree.

{\bf Acknowledgments:} DM thanks Seif Randjbar-Daemi
and the High Energy Group at Abdus Salam I.C.T.P. and
Jack Gunion and the High Energy 
Group at the Department of Physics, 
U.C. Davis for sabbatical leave hospitality while this work was being
completed. AJ and PJ thank S. J. Sciutto for help in using the 
AIRES air shower simulator. We also thank Jaime Alvarez for useful
discussions.  This work was supported in part by U.S. DOE Grant number
DE-FG03-98ER41079, 
the {\it Kansas Institute for Theoretical and Computational Science}
and DST (India) grant No. DST/PHY/19990184.


\begin{thebibliography}{unsrt}
\bibitem{usnote}Besides the $s^2$ behaviour 
several slow-growth models are also studied in \cite{us1}.
\end{thebibliography}
\end{document}